\documentstyle[11pt]{article}
\title{Volume of classical and quantum ensembles: geometric approach to
entropy and information}
\author{Michael J. W. Hall\\ \\Abteilung f\"{u}r Quantenphysik\\
Universit\"{a}t Ulm\\D-89069 Ulm, Germany}
\date{25 March 1998}
\begin{document}
\maketitle
\begin{abstract}
It is shown for classical and quantum ensembles 
that there is a unique
quantity which has the properties of a ``volume".  This quantity is a
function of the ensemble entropy, and hence provides a geometric 
interpretation for the latter.
It further provides a simple geometric picture for deriving and
unifying a number of results in classical and quantum information 
theory, and for discussing entropic uncertainty relations.

\end{abstract}
\newpage

Classical and quantum ensembles make natural appearances in
many different contexts.  These include, for example, equilibrium 
ensembles in statistical mechanics;  signal-state ensembles in 
communication theory;  dynamical ensembles corresponding to Brownian or
chaotic diffusion;  and ensembles 
corresponding to the classical limits of various quantum states.

In each of the above contexts it is often fruitful to employ the notion
of a ``volume" associated with the ensemble.  For example,
derivations in
statistical mechanics often involve counting ``microstates" 
in a volume of small thickness containing a constant-energy
surface \cite{statmech}.  Shannon's theorem for information transfer,
in the case of signals subject to quadratic energy and noise 
constraints, can be proved by 
considering the ratios of spherical volumes in
high-dimensional spaces \cite{shannon}.  In Ornstein-Uhlenbeck 
diffusion 
the evolution of a Gaussian ensemble is usefully depicted by 
a ``distribution ellipsoid" 
(with principal axis lengths corresponding to 
root-mean-square variances) 
\cite{diff}, where the ellipsoid volume is a clear measure of the 
``spread" of the ensemble.  Finally, {\it minimum} phase-space volumes
on the order of $h^{n}$ play a fundamental role in the classical
limit of quantum mechanics \cite{semiclass}.

The above examples raise the question of whether there is in fact some
{\it general} measure of ``volume" for classical and quantum ensembles,
which may be usefully employed in all of the above contexts but which
is not restricted in application or interpretation to various special
cases.  Thus, for example, the measure $\Delta x\Delta p$ for  
an ensemble of systems with a 2-dimensional phase space 
is not suitable, because (i) it is not invariant
under canonical/unitary transformations;  (ii) it has no natural
generalisation to finite-dimensional quantum systems; and (iii) 
while it
leads to useful geometric information bounds \cite{hallpra}, 
these are only exact in particular cases.

It will be seen here that there is indeed a natural 
measure of ``ensemble volume".
This measure is a function of the Gibb's entropy for classical ensembles
and of the von Neumann entropy for quantum ensembles, and hence provides
a unified geometric interpretation for these quantities.  

Before writing down this volume measure, it is useful
to consider four simple properties which uniquely define it
(up to a normalisation constant).  The first is an invariance property,
while the remaining three are geometric in nature.  
It is convenient to
state these properties in a form independent of whether the ensemble is
classical or quantum, as a common
geometric viewpoint is desirable.  
Thus, $\Gamma$ will be allowed to denote either a classical phase space
or a quantum Hilbert space; $\rho$ either a probability distribution on
phase space or a density operator on Hilbert space; and ${\rm
Tr}_{\Gamma}
[\cdot ]$ either integration over a classical phase space or
the trace over a quantum Hilbert space.  Moreover, $\Gamma_{12}$ will
denote the phase space or Hilbert space corresponding to an ensemble
of composite systems 
with subsystem 
spaces $\Gamma_{1}$ and $\Gamma_{2}$, i.e., $\Gamma_{12}$ denotes 
 $\Gamma_{1}\times\Gamma_{2}$
for classical ensembles and $\Gamma_{1}\otimes\Gamma_{2}$ for quantum
ensembles.  

Consider now a volume measure $V(\rho)$ which satisfies the
following properties:

{\it (i) Invariance Property:}  $V(\rho)$ is invariant under all 
canonical transformations (these are 
represented by unitary transformations
for quantum ensembles).  Note that this property ensures that the volume
is a function of the ensemble alone, independently of a particular 
co-ordinatisation or measurement basis.

{\it (ii) Cartesian Property:}  If $\rho$ describes two {\it
uncorrelated}
ensembles $\rho_{1}$ and $\rho_{2}$ on $\Gamma_{1}$ and $\Gamma_{2}$
respectively, then 
\begin{equation} \label{cart}
V(\rho) = V(\rho_{1}) V(\rho_{2}) .
\end{equation}
This property is exactly analogous to the geometric property that area
equals length times breadth, and is illustrated in Fig.~1.  Note that 
$\rho$ is the product $\rho_{1}$$\rho_{2}$ for classical ensembles, and
the tensor product $\rho_{1}$$\otimes$$\rho_{2}$ for quantum 
ensembles.

{\it (iii) Projection Property:}  If $\rho$ describes an ensemble of
composite systems on $\Gamma_{12}$ then
\begin{equation} \label{proj}
V(\rho) \leq V(\rho_{1}) V(\rho_{2}) ,
\end{equation}
where $\rho_{1}$, $\rho_{2}$ denote the ``projections" of $\rho$ onto
$\Gamma_{1}$, $\Gamma_{2}$ respectively (i.e., $\rho_{1}=$ 
${\rm Tr}_{\Gamma_{2}}[\rho]$, 
$\rho_{2}=$ ${\rm Tr}_{\Gamma_{1}}[\rho]$).  This
property is exactly analogous to the geometric property that a volume
is less than or equal to the product of the lengths obtained by
its projection onto orthogonal axes, and is illustrated in Fig.~2.
  Note that  $\rho_{1}$, $\rho_{2}$ denote the {\it marginal}
distributions corresponding to $\Gamma_{1}$, $\Gamma_{2}$ for classical
ensembles, and the {\it reduced} density operators corresponding to
 $\Gamma_{1}$, $\Gamma_{2}$ for quantum ensembles.

{\it (iv) Uniformity Property:}  If $\rho '$ and $\rho ''$ denote two
{\it non-overlapping} ensembles (i.e., 
${\rm Tr}_{\Gamma}[\rho '\rho '']$
$=0$), with equal volumes $V(\rho ')=$ $V(\rho '')=$ $V$, then an
arbitrary mixture $\rho$ of $\rho '$ and $\rho ''$ has volume no
greater than $2V$, where the latter corresponds to the volume of an
equally-weighted mixture, i.e.,
\begin{equation} \label{unif}
V(\rho) \leq V(\rho ' /2 + \rho '' /2) = 2 V .
\end{equation}
Thus uniform mixtures maximise ensemble volume.

One has the following result:

{\it Theorem:}  Any (continuous) measure of volume 
satisfying properties (i)-(iv)
above has the form
\begin{equation} \label{theo}
V(\rho) = K(\Gamma) e^{S(\rho)},
\end{equation}
where $S(\rho)$ denotes the ensemble entropy
\begin{equation} \label{ent}
S(\rho) = - {\rm Tr}_{\Gamma}[\rho \ln \rho] ,
\end{equation}
and $K({\Gamma})$ is a constant which may depend on $\Gamma$, and 
satisfies
\begin{equation} \label{kgam}
K(\Gamma_{12}) = K(\Gamma_{1}) K(\Gamma_{2}) .
\end{equation}

The proof will be given elsewhere for
reasons of space, and primarily
relies on applying properties (i)-(iv) to an arbitrarily large number
of independent copies of a given ensemble $\rho$ \cite{genpdf}.
To indicate its plausiblity here, consider
a quantum ensemble with eigenvalue distribution $\{\lambda_{j}\}$.  
Properties (i), (ii) and (iv) are certainly satisfied by the
``Renyi'' volumes
$(\sum_{j} (\lambda_{j})^{\alpha} )^{\beta}$ with
$(1-\alpha)\beta$ $=1$,
together with a term $K(\Gamma)$ satisfying Eq.~(\ref{kgam}).  
A result of Renyi \cite{renyi} implies then that the
projection property (iii) is satisfied only in the limit 
$\alpha\rightarrow 1$, yielding Eqs. (\ref{theo}) and (\ref{ent}).

The universal geometric interpretation of the 
volume measure in Eq.~(\ref{theo}) 
contrasts with ensemble entropy, for which the only
context-independent interpretation to date
appears to be as a somewhat vague measure of ``uncertainty"
or ``randomness" \cite{renyi,shan,maas}, 
which has a limited and purely
heuristic usefulness.  Indeed, it is perhaps conceptually
more useful to {\it define}
ensemble entropy as the logarithm of the ensemble volume $V(\rho)$,
thus providing a geometric picture for many of its properties.
Applications of the volume measure in Eq.~(\ref{theo}) to various
contexts will now be discussed.

First, in the statistical mechanics context, 
the Gibbs relation $S_{th}=k S(\rho)$ between thermodynamic entropy and
ensemble entropy for equilibrium ensembles can be rewritten via 
Eqs. (\ref{theo}) and (\ref{ent}) as
\begin{equation} \label{gibb}
S_{th} = k \ln [V(\rho)/K(\Gamma)] .
\end{equation}
Thus, the thermodyamic entropy is 
(up to an additive constant) 
proportional to the logarithm of the ensemble volume.  Note further 
from  Eq.~(\ref{gibb}) and the third law of thermodynamics 
(that thermodynamic entropy vanishes at absolute zero), that
$K(\Gamma)$ should correspond to 
a minimum ``zero-temperature" ensemble volume.  For quantum
ensembles one has from  Eqs. (\ref{theo}) and (\ref{ent}) that
$V(\rho)=K(\Gamma)$ for pure states, i.e., the 
{\it quantum} zero-temperature volume is just that of a 
{\it pure} state on $\Gamma$.  However, classical ensembles violate the
third law \cite{statmech} and  $K(\Gamma)$ remains arbitrary in this
case (but see Eq.~(\ref{corr}) below). 

The geometric expression (\ref{gibb}) is very similar 
to the original Boltzmann relation
$S_{th} = k \ln W$ ,
where $W$ is the number of distinct
microstates or ``elementary complexions" consistent with the
thermodynamic description.  Indeed, 
from the above discussion it follows that
Eq.~(\ref{gibb}) provides a {\it precise} 
{\it geometric} interpretation
of the Boltzmann relation for equilibrium ensembles:  thermodynamic 
entropy is proportional to the logarithm of  
the number of non-overlapping zero-temperature (or ``microstate")
volumes contained
within the total volume of the ensemble.

Second, in the communication context, consider a communication channel
where signal states $\rho_{1}$, $\rho_{2}$, $\dots$ are transmitted with
prior probabilities $p_{1}$, $p_{2}$, $\dots$ respectively
\cite{footsig}.  
The ensemble of signal
states itself corresponds to the mixture
\begin{equation} \label{ensem}
\rho = \sum_{i} p_{i} \rho_{i} .
\end{equation}
As a warm-up exercise, suppose that the 
signal states all have volumes greater than some minimum volume $V_{0}$
(e.g., due to noise in the channel).
From the uniformity property (iv) it follows that 
the maximum possible number of 
{\it non-overlapping} signal volumes available
for a given signal ensemble $\rho$
is bounded by $V(\rho)/V_{0}$.  Hence the maximum amount
of {\it error-free} data, $I_{1}$, which can be gained by a single
measurement at the receiver (measured in terms of the number of 
binary digits required to represent the data), is bounded by
\begin{equation} \label{single}
I_{1} \leq \log_{2} (V(\rho)/V_{0}) . 
\end{equation}

For example, if the channel is quantum 
one may always take $V_{0}=$ $K(\Gamma)$ (i.e., the volume of a
pure state).  Eqs. (\ref{theo}) and (\ref{single}) then immediately
yield the single-measurement quantum information bound
\begin{equation}
I_{1}^{Q} \leq S(\rho) \log_{2} e ,
\end{equation}
which may be recognised as a special case of Holevo's bound in
quantum communication theory \cite{hol} (see also below).

More generally, one may seek to improve information transfer by
coding data into {\it blocks} of signal states, of some length $L$, and 
restricting transmission to particular blocks of signals
\cite{shan}.  These
may be referred to as ``block-signals", to distinguish them from the
individual signals $\{\rho_{i}\}$.  The 
ensemble of block signals will be denoted by
$\rho^{(L)}$.  Under the constraint
that state $\rho_{i}$ still appears with relative frequency $p_{i}$
per individual signal transmission, one has
\begin{equation} \label{rhol}
V(\rho^{(L)}) \leq V(\overline{\rho}_{1})\dots V(\overline{\rho}_{L})
\leq [V(\rho)]^{L}  ,
\end{equation}
where $\overline{\rho}_{l}$ is the average $l$-th component of the
transmitted block-signals; the first inequality follows from the
projection property (iii); and the second inequality from the
concavity property 
\begin{equation} \label{conv}
\sum_{l} L^{-1} S(\overline{\rho}_{l}) \leq
S(\sum_{l} L^{-1} \overline{\rho}_{l}) = S(\rho) 
\nonumber \end{equation}
of entropy and Eqs. (\ref{theo}), (\ref{kgam}) and (\ref{ensem}).

Further, for $L$ sufficiently large, the
strong law of large numbers guarantees that most block-signals are
``typical" in the sense of Shannon \cite{shan}, i.e., 
the number of times that a signal state $\rho_{i}$ appears
in a given block-signal is approximately $p_{i}L$, except for a set of 
block-signals with a total probability of occurrence approaching zero as
$L$ is increased (and hence which may be ignored).  
It then follows immediately from  the Cartesian property~(\ref{cart})
that the volume of a typical block signal $\alpha$ is approximately
\begin{equation} \label{typvol}
V_{\alpha} = \prod_{i} [V(\rho_{i})]^{p_{i}L} 
\end{equation}
(and approaches it arbitrarily closely in the limit of large $L$).

Exactly as per the derivation of Eq.~(\ref{single}), 
the amount of error-free data $I$ which can be gained 
per individual transmitted signal, by measurements on block-signals, 
is geometrically bounded in the limit of
arbitrarily large block size by
\begin{equation} \label{block}
I \leq L^{-1} \log_{2} [V( \rho^{(L)})/ V_{\alpha}]  . 
\end{equation}
From Eqs. (\ref{theo}), (\ref{rhol}) and (\ref{typvol}) 
this yields the {\it general} upper bound
\begin{equation} \label{hol}
I \leq [S(\rho) - \sum_{i} p_{i} S(\rho_{i})] \log_{2} e  .
\end{equation}

For classical ensembles, Eq.~(\ref{hol}) may be recognised as the
{\it Shannon bound} for discrete memoryless channels \cite{shan}, 
while for quantum ensembles it may be recognised
as the {\it Holevo bound} for such
channels \cite{hol,cd,yuen}.  The unified derivation of these bounds
from simple volume considerations is a central result of this paper.
It is particularly valuable in the quantum case,
as existing derivations of the Holevo bound
are mathematically rather technical 
\cite{hol,yuen,fc}.  

It has recently been shown \cite{tight} that the Holevo bound
(\ref{hol}) is in fact {\it tight}, in the sense of being achievable 
arbitrarily closely for sufficiently large block sizes by choosing a
suitable set of typical block-signals 
for transmission and making appropriate 
measurements on these block-signals at the receiver.  Geometrically,
this result corresponds to being able to choose up to 
$[V(\rho)]^{L}/V_{\alpha}$ non-overlapping block-signal volumes,
in the limit of arbitrarily large $L$, and it would be valuable if a
geometrically-based proof of this could be given.  
Since the {\it total} number of typical
signal-blocks may be estimated as $\exp(-L\sum_{i} p_{i}\ln p_{i})$
\cite{shan}, tightness of bound Eq. (\ref{hol}) immediately implies
the Lanford-Robinson inequality \cite{wehrl}
\begin{equation} \label{lanf}
S(\rho) - \sum_{i} p_{i} S(\rho_{i}) \leq - \sum_{i} p_{i}\ln p_{i}  .
\end{equation}

Next, in the diffusion context, consider a Gaussian classical ensemble
with  root-mean-square variances $\Delta x_{1}$, $\dots$, $\Delta x_{n}$
, $\Delta p_{1}$, $\dots$, $\Delta p_{n}$ relative to the principal axes
(these variances and axes will in general
vary in time for Ornstein-Uhlenbeck
diffusion processes \cite{diff}).  The ensemble volume may then be
calculated from Eq.~(\ref{theo}) as
\begin{equation} \label{gauss}
V(\rho) = K(\Gamma) \prod_{j=1}^{n} 2\pi e \Delta x_{j}\Delta p_{j} ,
\end{equation}
which is directly proportional to the volume of the corresponding
``distribution  ellipsoid"  (with equality
for the choice $K(\Gamma)$ $=n!(2e)^{-n}$).  This volume is moreover
invariant under {\it all} canonical
transformations.  Note that trivially from Eq.~(\ref{theo})
irreversible dynamical processes are characterised by a strictly
increasing ensemble volume.

Consider finally a classical ensemble $\rho_{C}$ which is the 
``classical limit" of some quantum ensemble $\rho_{Q}$, i.e., the
physical properties of $\rho_{C}$ approximate those of 
$\rho_{Q}$.  For the case of a spinless particle on a 
$2n$-dimensional phase space one can obtain a relationship between the
constants $K(\Gamma_{C})$ and $K(\Gamma_{Q})$ in Eq.~(\ref{theo}) by
requiring that the ensemble volumes $V(\rho_{C})$ and $V(\rho_{Q})$
are approximately equal for such ensembles.  Since these constants are
independent of the dynamics of the ensemble it suffices to choose
an equilibrium ensemble of 
isotropic oscillators. 
Equating $V(\rho_{C})$ and $V(\rho_{Q})$ in the high-temperature
limit then yields
\begin{equation} \label{corr}
K(\Gamma_{Q}) = h^{n}  K(\Gamma_{C})  ,
\end{equation}
for the volume of a pure state, where $h$ is Planck's constant.

Eq.~(\ref{corr}) can be used to derive semi-classical uncertainty
relations from geometric considerations.  For two
corresponding ensembles  $\rho_{Q}$ and  $\rho_{C}$ as above 
the position and momentum entropies $S(\rho_{X})$ and $S(\rho_{P})$
will be approximately equivalent for either ensemble, and further
\begin{equation} \label{uncert}
\exp(S(\rho_{X})) \exp(S(\rho_{P})) \geq \exp(S(\rho_{C})
\end{equation}
holds for the classical ensemble from the projection property (ii)
applied to projections onto the position and momentum axes rather
than spaces $\Gamma_{1}$ and $\Gamma_{2}$ \cite{proj}.  
Eqs. (\ref{theo}), (\ref{corr}) and 
(\ref{uncert}) yield
\begin{equation} \label{entrop}
S(\rho_{X}) + S(\rho_{P}) \stackrel{>}{\sim} n \ln h  .
\end{equation}
This semi-classical entropic 
uncertainty relation,
holding for quantum ensembles which have classical limits, 
arises directly from the existence of a {\it minimum} volume 
for quantum ensembles.  Following the method of \cite{bbm}, the
corresponding semi-classical Heisenberg uncertainty relation
\begin{equation} \label{heis}
\Delta x \Delta p \stackrel{>}{\sim} \hbar/e  
\end{equation}
follows for the $n=1$ case.  Eqs. (\ref{entrop}) and (\ref{heis}) are
close to the exact results for
general quantum ensembles \cite{bbm} (where, e.g., $e$ is 
replaced by $2$ in Eq.~(\ref{heis})). 

In conclusion, an essentially unique measure of volume for classical
and quantum ensembles has been found, related to ensemble entropy,
which provides a 
geometric tool for any context in which ensembles appear.
This measure is  universal in the sense that it may
be defined by theory-independent concepts of invariance, 
uncorrelated ensembles, projection, and non-overlapping ensembles
(properties (i)-(iv)).  Applications include a precise geometric
interpretation of the Boltzmann relation; a unified
derivation of results in classical and quantum information theory
based on simple geometric properties; an invariant generalisation of
``ellipsoid volume" for diffusion problems; and a geometric
derivation of semi-classical uncertainty relations.

This work was supported by the Alexander von Humboldt Foundation.

\newpage
{\bf FIGURE CAPTIONS}
\\
\\
\\
FIG. 1.  Two uncorrelated ensembles $\rho_{1}$ and $\rho_{2}$ on
spaces $\Gamma_{1}$ and $\Gamma_{2}$ respectively (shown here
compressed to 1-dimensional axes), have respective volumes
$V(\rho_{1})$ and $V(\rho_{2})$ as indicated by the darkened axis 
regions.  The {\it Cartesian property} Eq.~(1) states that the
corresponding joint ensemble $\rho$ has a ``rectangular"
volume $V(\rho)=V(\rho_{1})V(\rho_{2})$, i.e., $V(\rho)$
corresponds to the Cartesian product of volumes $V(\rho_{1})$
and $V(\rho_{2})$.
\\
\\
\\
FIG. 2.  An ensemble $\rho$ on the product space of  $\Gamma_{1}$
and $\Gamma_{2}$ has a volume $V(\rho)$ indicated by the 
solid closed curve.  The corresponding projected (or ``reduced") 
ensembles  $\rho_{1}$ and $\rho_{2}$ on  $\Gamma_{1}$ and $\Gamma_{2}$
respectively have projected volumes  $V(\rho_{1})$ and
 $V(\rho_{2})$, indicated by the darkened axis regions.  The
{\it projection property} Eq.~(2) states that $V(\rho)$ can be no
greater than the volume of the rectangular region formed by the 
dashed lines, i.e., than the product of the projected
volumes.

\newpage
\begin{thebibliography}{99}
\bibitem[1]{statmech}  M. Toda, R.Kubo and
N. Sait\^o {\it Statistical Physics I} (Springer, Berlin, 1983), 
Sec. 2.1.  
\bibitem[2]{shannon} C.E Shannon, Proc. IRE {\bf 37}, 160 (1949),
reprinted in {\it Claude Elwood Shannon: Collected Papers}, edited by
N. Sloane and A. Wyner (IEEE, New York, 1993), pp. 160-172.
\bibitem[3]{diff} M.C. Wang and G.E. Uhlenbeck, Rev. Mod. Phys. 
{\bf 17}, 323 (1945), Sec. 10.
\bibitem[4]{semiclass} A.S. Davydov, {\it Quantum Mechanics} 2nd edn.
(Pergamon Press, UK, 1976), Sec. III.23.
\bibitem[5]{hallpra} M.J.W. Hall, Phys. Rev. A {\bf 55}, 100 (1997).
\bibitem[6]{genpdf} The theorem may 
be generalised to discrete and/or
non-dynamical classical ensembles, by rewriting the invariance
property (i) as 
the requirement that $V(\rho)$ is invariant under
${\rm Tr}_{\Gamma}[\cdot ]$-preserving transformations.
\bibitem[7]{renyi} A. Renyi, {\it Probability Theory} (North-Holland,
Amsterdam, 1970), Sec. IX.6, theorem 4.
\bibitem[8]{shan} C.E. Shannon, Bell Syst. Tech J. {\bf 27}, 379
(1948); {\bf 27}, 623, reprinted in the collection of Ref. [2] above,
pp. 5-83.
\bibitem[9]{maas} H. Maassen and J.B.M. Uffink, Phys. Rev. Lett.
{\bf 60}, 1103 (1988).
\bibitem[10]{footsig} The signal states are in general represented 
by ensembles, to take into
account any noise processes in the transmitter and channel medium prior
to detection at the receiver.
\bibitem[11]{hol} A.S. Holevo, Probl.
Inf. Trans. {\bf 9}, 177 (1973).
\bibitem[12]{cd} C.M. Caves and P.D. Drummond, Rev. Mod. Phys.
{\bf 66}, 481 (1994).
\bibitem[13]{yuen} H.P. Yuen and M. Ozawa, Phys. Rev. Lett. {\bf 70},
363 (1993).
\bibitem[14]{fc} C.A. Fuchs and C.M. Caves, Phys. Rev. Lett. {\bf 73},
3047 (1994).
\bibitem[15]{tight} B. Schumacher and M.D. Westmoreland, Phys. Rev. A
{\bf 56}, 131 (1997);  A.S. Holevo, quant-ph/9611023.
\bibitem[16]{wehrl} A. Wehrl, Rev. Mod. Phys. {\bf 50}, 221 (1978).
\bibitem[17]{proj} Note this provides a geometric interpretation for
the property that the entropy of a joint distribution is not greater
than the sum of the entropies of its marginal distributions (see,
e.g., Secs. 6, 20 of \cite{shan}).
\bibitem[18]{bbm} I. Bialynicki-Birula and J. Mycielski, Commun. Math.
Phys. {\bf 44}, 129 (1975).
\end{thebibliography}
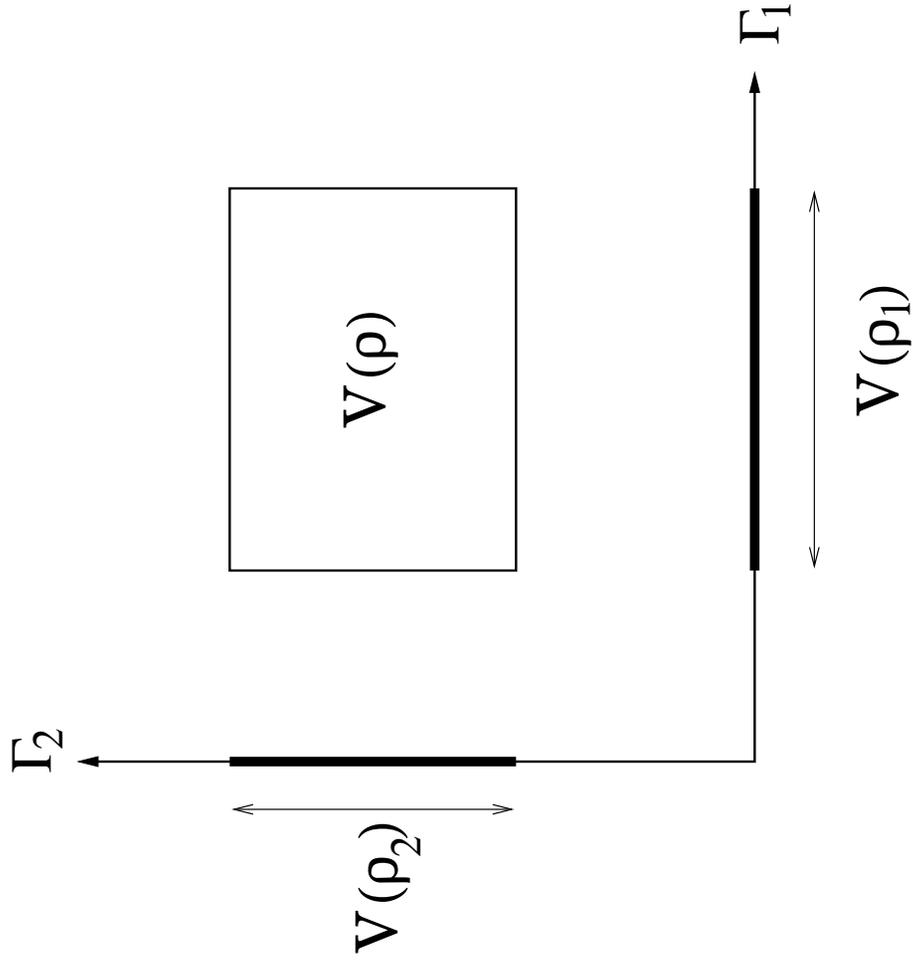
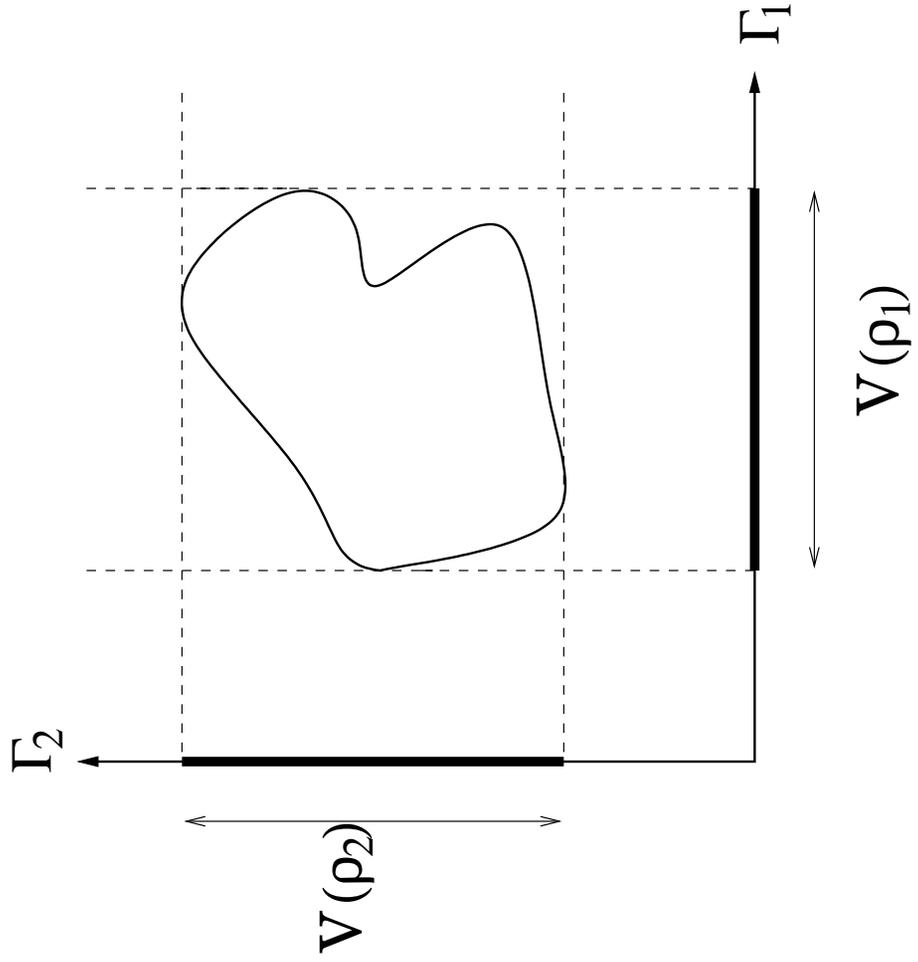
\end{document}